\documentclass[12pt]{article}

\usepackage{epsfig}

\newcommand{\newsection}{\section}

\def\be{\begin{equation}}
\def\ee{\end{equation}}
\newcommand{\bea}{\begin{eqnarray}}
\newcommand{\eea}{\end{eqnarray}}
\def\bd{\begin{displaymath}}
\def\ed{\end{displaymath}}

\def\ba{\begin{eqnarray}}
\def\ea{\end{eqnarray}}

\newcommand{\non}{\nonumber \\}
\newcommand{\CR}{\non\cr}

\begin{document}

\vspace{18pt}
\par\hfill WIS/05/06-APR-DPP
\vskip 0.01in \hfill TAUP-2824/06
\vskip 0.01in \hfill {\tt hep-th/0604161}

\vspace{30pt}

\begin{center}
{\bf \LARGE A Holographic Model of Deconfinement and}
\end{center}
\begin{center}
{\bf \LARGE Chiral Symmetry
Restoration}
\end{center}

\vspace{30pt}

\begin{center}
Ofer Aharony$\,^a$, Jacob Sonnenschein$\,^{b,c}$, and Shimon Yankielowicz$\,^{b,c}$

\vspace{20pt}

\textit{$^a$Department of Particle Physics\\ The Weizmann Institute of 
Science, Rehovot 76100, Israel\\[10pt]
$^b$School of Physics and Astronomy\\ The Raymond and Beverly Sackler 
Faculty of Exact Sciences\\ Tel Aviv University, Ramat Aviv 69978,
Israel\\[10pt]
$^c$Albert Einstein Minerva Center\\ The Weizmann Institute of
Science, Rehovot 76100, Israel
}

\end{center}


\begin{center}
\textbf{Abstract }
\end{center}
We analyze the finite temperature behavior of the Sakai-Sugimoto
model, which is a holographic dual of a theory which spontaneously
breaks a $U(N_f)_L\times U(N_f)_R$ chiral flavor symmetry at zero
temperature. The theory involved is a $4+1$ dimensional supersymmetric
$SU(N_c)$ gauge theory compactified on a circle of radius $R$ with
anti-periodic boundary conditions for fermions, coupled to $N_f$
left-handed quarks and $N_f$ right-handed quarks which are localized
at different points on the compact circle (separated by a distance
$L$). In the supergravity limit which we analyze (corresponding in
particular to the large $N_c$ limit of the gauge theory), the theory
undergoes a deconfinement phase transition at a temperature $T_d = 1 /
2\pi R$. For quark separations obeying $L > L_c \simeq 0.97*R$ the
chiral symmetry is restored at this temperature, but for $L < L_c
\simeq 0.97*R$ there is an intermediate phase which is deconfined with
broken chiral symmetry, and the chiral symmetry is restored at
$T_{\chi SB} \simeq 0.154 / L$.  All of these phase transitions are of
first order.

\vspace{4pt} {\small \noindent 

 }  \vfill
\vskip 5.mm
 \hrule width 5.cm
\vskip 2.mm
{\small
\noindent
E-mails : Ofer.Aharony@weizmann.ac.il, cobi@post.tau.ac.il, 
shimonya@post.tau.ac.il.}

\thispagestyle{empty}

\eject

\setcounter{page}{1}

\newsection{Introduction}

QCD, the theory of the strong nuclear interactions, exhibits two
interesting phenomena at low energies -- confinement and (approximate)
spontaneous
chiral symmetry breaking. Both of these phenomena are strong coupling
effects which are not visible in perturbation theory, and there are no
known quantitative methods to study them (except by lattice simulations).
Apriori there is no relation between these two phenomena, and there are
known examples of chiral symmetry breaking in non-confining theories
and of confinement without chiral symmetry breaking. However, in QCD
both phenomena occur together.

New methods for studying strongly coupled gauge theories have been
developed in the last decade, following the AdS/CFT correspondence
\cite{Maldacena:1997re}. These methods are particularly powerful in large $N_c$ gauge
theories which are dual to weakly coupled string theories in weakly
curved spaces, since then many computations can be explicitly
performed. This class of theories does not include (large $N_c$) QCD,
but there are several known cases of theories which are labeled by a
dimensionless parameter, such that in one limit of the parameter the
string background is weakly curved, while the other limit gives (large
$N_c$) QCD \cite{Wpure,PS,KS,MN}. It is believed that these theories
are (for any value of the dimensionless parameter) in the same
universality class as QCD, so it is interesting to try to use these
theories to understand better the phenomena of confinement and chiral
symmetry breaking.

Understanding confinement in the weakly curved limit is amazingly
easy; the confining strings are explicitly seen as strings in the
weakly curved background (either fundamental strings or strings
arising from D-branes), and their tensions are easily computable. It
would be nice to be able to similarly study the phenomenon of chiral
symmetry breaking ($\chi$SB), and to see if (in the weakly curved
limit) there is any link between $\chi$SB and
confinement. Unfortunately, it seems much more difficult to find dual
models for theories with non-Abelian $U(N_f)_L\times U(N_f)_R$ chiral
symmetry than to find dual models for confining theories.
In the context of holographic dual models, quarks in the fundamental
representation of $SU(N_c)$ were first incorporated in terms of probe
D branes in \cite{Karch}\footnote{The first appearance of quarks in
the fundamental representation of $Sp(N_c)$ was in
\cite{Aharony:1998xz}.}.  Since then several holographic models dual
to theories with quarks have been constructed and investigated, with the
quarks coming from additional ``flavor D-branes'', in
both conformal and confining gauge theories \cite{SakSon,
Babington:2003vm,KMMW2,problist,KPSV,Bak:2004nt,CPS,PSZ}.  
However, there are relatively few brane
constructions with non-Abelian chiral symmetry (see, e.g.,
\cite{Brodie:1997sz}), and it has proven difficult to use these
brane constructions to construct (by taking a near-horizon limit)
string theory duals of theories with non-Abelian chiral symmetry (with
or without $\chi$SB)\footnote{On the other hand, there are several
models realizing a $U(1)$ axial flavor symmetry, such as
\cite{Babington:2003vm,KMMW2}.}.

This situation was recently changed by the work of Sakai and Sugimoto
\cite{SS,SS2}, who found a simple brane construction realizing chiral
symmetry, and took its near-horizon limit to find a string theory dual
to a field theory with chiral symmetry breaking. The brane
construction is in type IIA string theory compactified on a circle of
radius $R$ in the $x_4$ direction, and involves $N_c$ D4-branes
wrapped on the circle (filling the $01234$ directions), $N_f$
D8-branes sitting at the point $x_4=0$ and $N_f$ anti-D8-branes
sitting at the point $x_4=L$ \footnote{In \cite{SS} the specific
choice $L=\pi R$ was made, for which the D8's and anti-D8's are
antipodal on the circle, but this is not really necessary.}. The
circle could have either periodic or anti-periodic boundary conditions
for the fermions; the more interesting case for relating to QCD is the
case in which all fermions have anti-periodic boundary conditions
around the circle.  The gauge theory on the D4-branes at low energies
is a $U(N_c)$ $4+1$ dimensional maximally supersymmetric Yang-Mills
theory, which is coupled to $N_f$ left-handed $3+1$-dimensional chiral
fermions (in the fundamental representation of $U(N_c)$) at the
intersection with the D8-branes and to $N_f$ right-handed chiral
fermions at the intersection with the anti-D8-branes. With
anti-periodic boundary conditions, the fermions in the $4+1$
dimensional gauge multiplet obtain a mass of order $1/R$, and the
scalars obtain masses from quantum corrections, so the (classical) massless
spectrum is that of a $3+1$ dimensional $U(N_c)$ gauge theory coupled
to $N_f$ massless Dirac fermions.  The $U(N_f)_L\times U(N_f)_R$
global chiral symmetry of this theory\footnote{The axial $U(1)$ is
anomalous, but this is a subleading effect in the large $N_c$ limit
which will not be relevant for our leading order computations.} is
explicitly visible as the gauge symmetry on the D8 and anti-D8-branes.

The string theory described above is non-supersymmetric and
unstable, but this instability does not
affect the low-energy behavior, and disappears when we take a
decoupling limit for the theory on the D4-branes;
the unstable modes, which are
the scalar fields corresponding to the radius of the circle and to the
asymptotic brane-anti-brane separation,
become non-normalizable in this near-horizon limit so we do not
need to worry about them.
This decoupling
limit, before the compactification on a circle, was described in
\cite{Itzhaki:1998dd}, and it involves taking the string scale
$\alpha'$ to zero and the asymptotic
string coupling $g_s$ to infinity while
keeping fixed the $4+1$ dimensional gauge coupling $g_5^2 \sim g_s
\sqrt{\alpha'}$. The fact that the (asymptotic) string coupling goes
to infinity in this limit is related to the fact that the $4+1$
dimensional gauge theory is not well-defined at high energies, and
should be defined with some UV completion; in type IIA string theory
the completion defines this theory as the $5+1$-dimensional ${\cal
N}=(2,0)$ superconformal theory of M5-branes compactified on a circle,
but this will not be relevant for our purposes (since we will always
be interested in energy scales well below $1/g_5^2$). The near-horizon
limit of the D4-branes, which is dual to their (UV-completed)
$4+1$-dimensional supersymmetric gauge theory, was described in
\cite{Itzhaki:1998dd}. It is highly curved at small radial coordinates
(corresponding to the fact that the gauge theory is IR-free), and at
very large radial coordinates the string coupling becomes large and
the background needs to be lifted to M theory, but for large $N_c$
there is a large range of values of the radial coordinate for which
type IIA string theory gives a good description.

The near-horizon limit of D4-branes compactified on a circle with
anti-periodic boundary conditions was first discussed in \cite{Wpure}, where it
was noted that it was the same as the (Wick rotation of the)
near-horizon limit of near-extremal D4-branes, and that when the
radius of the circle is small compared to the scale of the 't Hooft
coupling of the Yang-Mills
theory, the background becomes weakly curved everywhere. 
This was, in fact, the first model of a confining theory
constructed following the AdS/CFT correspondence. In the limit of
large $N_c$ with fixed $N_f$, the back-reaction of the D8-branes is
small, and they can be treated as probes in this background. Sakai and
Sugimoto noted that in this setup there is a very nice
interpretation of chiral symmetry breaking\footnote{A similar 
  structure of probe branes in a confining background
  was previously introduced in \cite{SakSon} 
in the context of the model of \cite{KS}.}. 
At large radial positions
(corresponding to the high-energy limit of the field theory) the D8-branes and
anti-D8-branes are separated and one sees the full $U(N_f)_L\times
U(N_f)_R$ chiral symmetry. Apriori, as we go into the interior of
space, the D8-branes and the anti-D8-branes
could remain separated, or they could join
together into $N_f$ continuous 8-branes, so that only a single
(diagonal) $U(N_f)$ symmetry is visible, corresponding to spontaneous chiral
symmetry breaking. In the background of \cite{Wpure}, the radial position has a
minimal value (where the $x_4$-circle shrinks to zero size) and the
8-branes have nowhere to end, and therefore the D8-branes have to
smoothly join the anti-D8-branes and exhibit $\chi$SB.  In \cite{SS} an
explicit solution for the position of the 8-branes was constructed (in
the probe approximation) which realizes this. 

Recently, it was realized
in \cite{AHJK} that the scale of chiral symmetry breaking in this model is
actually independent of the scale of confinement (the first depends
mostly on $L$ while the second depends mostly on $R$), and that chiral
symmetry breaking persists even in the $R\to \infty$ limit in which
confinement disappears\footnote{The independence of the two phenomena in
the gravity approximation was first noted for an Abelian chiral symmetry
in \cite{Bak:2004nt}.}. We will see this separation of scales
also in our finite temperature analysis.

In this paper we will study the model of Sakai and Sugimoto at finite
temperature.  In asymptotically free gauge theories with confinement
and chiral symmetry breaking, both of these properties disappear at
high temperatures as the effective coupling becomes small. In QCD
there is no sharp phase transition distinguishing the low-temperature
and high-temperature phases, since there are no sharp order parameters
for confinement (due to the presence of dynamical quarks) and for
the (approximate) chiral symmetry
breaking. However, in a theory with exactly massless quarks one
expects to have a sharp phase transition associated with the chiral
symmetry restoration (characterized by the chiral condensate as an
order parameter), and in the large $N_c$ limit the deconfinement
transition also becomes a sharp phase transition (since the quarks
can be ignored in this limit, and the pure gauge theory at finite
temperature has a $Z_{N_c}$ global symmetry which is unbroken in the
confined phase and broken in the deconfined phase).  It is interesting
to ask whether these two phase transitions occur at the same temperature or
not. In the large $N_c$ limit of QCD this is a difficult question to
answer analytically. However, in the Sakai-Sugimoto model (which is
continuously related to large $N_c$ QCD by varying parameters) we will
see that the question is easy to answer\footnote{Note that this theory
is not asymptotically free, so apriori it is not clear whether at high
temperatures it should deconfine and restore chiral symmetry or
not.}. We will find that simple geometrical arguments imply that
chiral symmetry is always broken in the confined phase, so that chiral
symmetry restoration can happen either at the same temperature as
deconfinement or above it. A more detailed analysis will be necessary
to decide between these two possibilities, and we will find that in
the weakly-curved limit the answer to this question depends on the ratio
$L/R$ (which can go between zero and $\pi$; in the limit of small $L/R$
this ratio is inversely related to the ratio between the low-spin meson
masses and the low-spin glueball masses). We will show that for
$L/R > 0.97$ the two phase transitions occur together, while for
$L/R < 0.97$ the
chiral symmetry restoration transition
happens at a higher temperature than the deconfinement
transition. In the supergravity limit all of these phase
transitions are of first order.

We begin in section 2 with a description of the system at zero temperature.
In particular we present  
a short review of the Sakai-Sugimoto model and its meson spectrum,
generalizing the construction of \cite{SS} to arbitrary values
of $L/R$. In section 3 we discuss the finite temperature behavior of
this theory. We start by discussing the deconfinement transition, which
is associated with the bulk physics, and its implications on chiral
symmetry breaking. We continue by studying the realization of chiral
symmetry in the deconfined phase and computing the chiral symmetry
restoration temperature. We also discuss the spectrum of quark-anti-quark 
excitations
of the theory at finite temperature. We end in section 4 with our
conclusions. An appendix contains the computation of the difference
of free energies which is relevant for the deconfinement transition.


\newsection{The zero temperature string theory model}

The string theory dual of the gauge theory we described above at zero
temperature is the Sakai-Sugimoto model \cite{SS}. We start by
reviewing the basic structure of this model. We then describe the
spectrum of mesons of low and high spins.

\subsection{Review of the Sakai-Sugimoto model}

The model of Sakai and Sugimoto \cite{SS} describes the near-horizon
limit of a configuration of $N_c$ D4-branes, wrapping a circle in the
$x_4$ direction with anti-periodic  boundary conditions for the fermions,
and intersected by $N_f$
D8-branes (located at $x_4=0$) and $N_f$ anti-D8-branes (located at
$x_4=L$). This is dual to a $4+1$ dimensional maximally
supersymmetric $SU(N_c)$ Yang-Mills theory (with coupling constant $g_5$ and
with a specific UV completion that will not be important for us),
compactified on a circle of radius $R$ with anti-periodic boundary conditions
for the fermions, with $N_f$ left-handed quarks located
at $x_4=0$ and $N_f$ right-handed quarks located at $x_4=L$ (obviously we
can assume $L \leq \pi R$).

In the limit $N_f \ll N_c$, this background can be described by
$N_f$ probe D8-branes inserted into the
near-horizon limit of a set of $N_c$
D4-branes compactified on a circle with anti-periodic boundary conditions
for the fermions \cite{Wpure}. This background is simply related to the
(near-horizon limit of the) background of near-extremal D4-branes by
exchanging the roles (and signatures) of the time direction and of one
of the spatial directions.
Let us now briefly review this
model, emphasizing the manifestations of confinement and chiral
symmetry breaking.  The background of type IIA string theory
is characterized by the metric, the
RR four-form and a dilaton given by 
\bea\label{SSmodel} ds^2&=&\left(
\frac{u}{R_{D4}} \right)^{3/2}\left [- dt^2 +\delta_{ij}dx^i dx^j +
f(u) dx_4^2 \right ] +\left( \frac{R_{D4}}{u} \right)^{3/2} \left [
\frac{du^2}{f(u)} + u^2 d\Omega_4^2 \right ],
\CR
F_{(4)}&=& \frac{2\pi N_c}{V_4}\epsilon_4, \quad  
e^\phi = g_s\left( \frac{u}{R_{D4}} \right)^{3/4}, 
\quad
 R_{D4}^3 \equiv \pi g_s N_c l_s^3,\quad 
f(u)\equiv 1-\left( \frac{u_\Lambda}{u} \right)^3,
\eea   
where $t$ is the time direction and $x^i$ ($i=1,2,3$) are the
 uncompactified world-volume coordinates of the D4 branes, $x_4$ is a
 compactified direction of the D4-brane world-volume which is transverse to
 the probe D8 branes, the volume of the unit four-sphere $\Omega_4$ is
 denoted by $V_4$ and the corresponding volume form by $\epsilon_4$,
 $l_s$ is the string length and finally $g_s$ is a parameter related
 to the string coupling.  The submanifold of the background spanned by
 $x_4$ and $u$ has the topology of a cigar (as on the left-hand side of
 figure \ref{phases} below) where the minimum value of $u$ at
 the
 tip of the
 cigar is $u_\Lambda$. The tip of the cigar is non-singular if and
 only if the periodicity of $x_4$ is
\be 
\label{relru}
\delta
 x_4 = \frac{4\pi}{3}\left( \frac{R_{D4}^3}{u_\Lambda} \right)^{1/2} = 2\pi R
 \ee 
and we identify this with the periodicity of the circle that the
$4+1$-dimensional gauge theory lives on.

The parameters of this gauge theory, the five-dimensional gauge coupling
$g_5$, the low-energy four-dimensional 
 gauge coupling $g_4$, the glueball mass scale $M_{gb}$, and the
 string tension $T_{st}$ are determined from the background
 (\ref{SSmodel}) in
 the following form \cite{KMMW2} :
\bea\label{stringauge}
g_5^2&=&(2\pi)^2 g_s l_s,\qquad 
g^2_{4}=\frac{g_5^2}{2\pi R}=
3\sqrt{\pi}\left ( \frac{g_s u_\Lambda}{N_c l_s}\right )^{1/2},
 \qquad 
M_{gb} = \frac{1}{R},
\CR T_{st} &=& \frac{1}{2\pi
 l_s^2}\sqrt{g_{tt}g_{xx}}|_{u=u_\Lambda}= \frac{1}{2\pi l_s^2}\left(
 \frac{u_\Lambda}{R_{D4}} \right)^{3/2} =\frac{2}{27\pi} \frac{g^2_4 N_c}{R^2} 
= \frac{\lambda_5}{27\pi^2 R^3},\eea
where $\lambda_5 \equiv g_5^2 N_c$, $M_{gb}$ is the typical scale of
the glueball masses computed from the spectrum of excitations around
(\ref{SSmodel}), and $T_{st}$ is the confining string tension in this
model (given by the tension of a fundamental string stretched at $u=u_{\Lambda}$
where its energy is minimized). The gravity approximation is valid
whenever $\lambda_5 \gg R$, otherwise the curvature at $u \sim
u_{\Lambda}$ becomes large. Note that as usual in gravity
approximations of confining gauge theories, the string tension is much
larger than the glueball mass scale in this limit. At very large values of $u$ the
dilaton becomes large, but this happens at values which are of order
$N_c^{4/3}$ (in the
large $N_c$ limit with fixed $\lambda_5$), so this will
play no role in the large $N_c$ limit that we will be interested in.
The Wilson line of this gauge theory (before putting in the D8-branes)
admits an area law behavior \cite{BISY2}, as can be easily seen using
the conditions for confinement of \cite{KSS}.

Naively, at energies lower than the Kaluza-Klein scale $1 / R$ the
 dual gauge theory is effectively four dimensional; however, it turns out
that the theory confines and develops a mass gap of order $M_{gb}=1/R$, so (in the
regime where the gravity approximation is valid) there
is no real separation between the confined four-dimensional fields and
the higher Kaluza-Klein modes on the circle. As discussed in \cite{Wpure},
in the opposite limit of $\lambda_5 \ll R$, the theory approaches the
$3+1$ dimensional pure Yang-Mills theory at energies small compared to
$1/R$, since in this limit the scale of the mass gap is exponentially
small compared to $1/R$. It is believed that there is no phase transition
when varying $\lambda_5/R$ between the gravity regime and the pure
 Yang-Mills regime, but it is not clear how to check this.

Next, we introduce
the probe 8-branes which
span the coordinates $t, x^i, \Omega_4$, and follow some curve $u(x_4)$
in the $(x_4,u)$-plane. Near the boundary at $u\to \infty$ we want to have
$N_f$ D8-branes localized at $x_4=0$ and $N_f$ anti-D8-branes (or D8-branes
with an opposite orientation) localized at $x_4=L$. Naively one might think
that the D8-branes and anti-D8-branes would go into the interior of the space
and stay disconnected; however, these 8-branes
do not have anywhere to end in the background (\ref{SSmodel}), so the
form of $u(x_4)$ must be such that the D8-branes smoothly connect to the 
anti-D8-branes (namely, $u$ must go to infinity at $x_4=0$ and at $x_4=L$,
and $du/dx_4$ must vanish at some minimal $u$ coordinate $u=u_0$). Such
a configuration spontaneously
breaks the chiral symmetry from the symmetry group which is
visible at large $u$, 
$U(N_f)_L\times U(N_f)_R$, to the diagonal $U(N_f)$ symmetry.
Thus, in this configuration the topology forces a breaking of the chiral
symmetry; this is not too surprising since chiral symmetry breaking at
large $N_c$ follows from rather simple considerations \cite{Coleman:1980mx}.

In order to find the 8-brane configuration, we need
the induced metric on the D8-branes, which is
\bea
\label{induced}
ds^2_{D8}&=&\left( \frac{u}{R_{D4}} \right)^{3/2}\left [ - dt^2+  \delta_{ij}dx^{i}dx^j \right ] 
+\left( \frac{u}{R_{D4}} \right)^{3/2} \left [f(u) + \left( \frac{R_{D4}}{u} \right)^{3} \frac{{u'}^2}{f(u)}\right ]dx_4^2 \CR
 &+&
 \left( \frac{R_{D4}}{u} \right)^{3/2} u^2 d\Omega_4^2  
\eea  
where $u'=du/dx_4$.
It is easy to check that the CS term in the D8-brane action does not affect the
solution of the equations of motion.  More precisely, the equation
of motion of the gauge field has a classical solution of a vanishing
gauge field, since the CS term includes terms of the form $C_5\wedge
F\wedge F$ and $C_3 \wedge F\wedge F\wedge F$. So, we are left only with
the DBI action.  Substituting the determinant of the induced metric
and the dilaton into the DBI action, we obtain
(ignoring the factor of $N_f$ which multiplies all the D8-brane
actions that we will write) :
\bea
\label{eightaction}
S_{DBI} =  T_8 \int dt d^3 x d x_4 d^4\Omega e^{-\phi} 
\sqrt{-\det(\hat g)}
        =  \frac{\hat T_8 }{g_s}\int  dx_4 u^4 \sqrt{f(u)+ 
\left( \frac{R_{D4}}{u} \right)^{3} \frac{{u'}^2}{f(u)}},
\eea 
where $\hat g$ is the induced metric (\ref{induced}) and $\hat T_8$ includes the
 outcome of the integration over all the coordinates apart from $d
 x_4$.

\begin{figure}[t]
\begin{center}
\vspace{3ex}
\includegraphics[width=.65\textwidth]{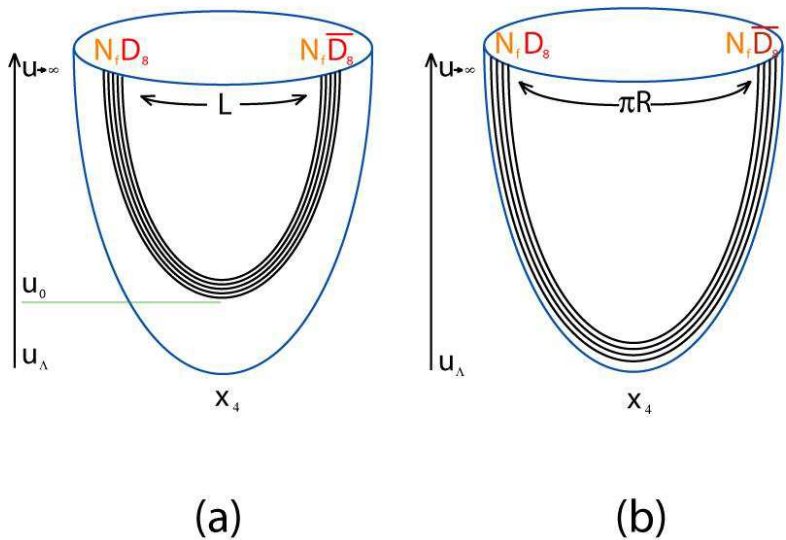}
\end{center}
\caption{The dominant configurations of the D8 and anti-D8 probe
branes in the Sakai-Sugimoto model at zero temperature, 
which break the chiral symmetry. The same configurations will turn
out to be relevant also at low temperatures.
On the left a generic configuration with an asymptotic separation of
$L$, that stretches down to a minimum at $u=u_0$, is drawn. The figure
on the right describes the limiting antipodal case $L=\pi R$, where
the branes connect at $u_0=u_{\Lambda}$. \label{D8-barD8SS}}
\end{figure}

 The simplest way to solve the equation of motion is by using the
 Hamiltonian of the action (\ref{eightaction}), which is conserved
 (independent of $x_4$) :
\be\label{D8profile} \frac{u^4 f(u)}{\sqrt{f(u)+ \left(
\frac{R_{D4}}{u} \right)^{3} \frac{{u'}^2}{f(u)}}}= {\rm constant} =
u_0^4 \sqrt{f(u_0)}, \ee 
where on the right-hand side of the equation we assumed
that there is a point $u_0$ where the profile $u(x_4)$ of the brane has a
minimum, $u'(u=u_0)=0$\footnote{This type of analysis was done
  previously for
  Wilson line configurations. See, for instance,
  \cite{BISY2}.}.
We need to find the solution in which 
as $u$ goes to infinity, $x_4$ goes to the values $x_4=0,L$; this
implies
\be
\int dx_4 = 2 \int {du \over u'} = L 
\label{forl}
\ee 
with $u'$ given (as a function of $u$) by
(\ref{D8profile}) (note that $u$ is a double-valued function of $x_4$
in these configurations, leading to the factor of two in (\ref{forl})).
The form of this profile of the D8 brane is drawn in figure
\ref{D8-barD8SS}(a). Plugging in the value of $u'$ from (\ref{D8profile})
we find
\bea
\label{lzerotemp}
L&=&\int dx_4 = 2\int_{u_0}^{\infty} \frac{du}{u'}= 
2 R^{3/2}_{D4}
\int_{u_0}^\infty du \frac{1}{f(u)u^{3/2}\sqrt{ \frac{f(u) u^8}{f(u_0) u_0^8}
-1}} \CR
&=&  \frac{2}{3} \left( \frac{R^3_{D4}}{u_0} \right)^{1/2}
\sqrt{1-y_\Lambda^3}\int_0^1 dz \frac{z^{1/2}}{(1-y_\Lambda^3 z)
\sqrt{1-y_\Lambda^3 z -(1-y_\Lambda^3)z^{8/3}}},
\eea
where $y_\Lambda \equiv u_{\Lambda} / u_0$.
Small values of $L$ correspond to large values of $u_0$. In this limit
we have $y_{\Lambda}\ll 1$ leading to $L\propto \sqrt{R_{D4}^3/u_0}$.
For general values of $L$ the dependence of $u_0$ on $L$ is more complicated.

There is a simple special case of the above solutions, which occurs
when $L = \pi R$, namely the D8-branes and anti-D8-branes lie at
antipodal points of the circle. In this case the solution for the
branes is simply $x_4(u) = 0$ and $x_4(u)=L=\pi R$, with the two branches
meeting smoothly at the minimal value $u=u_0=u_{\Lambda}$ to join
the D8-branes and the anti-D8-branes together. 
This type of antipodal solution is drawn in figure
\ref{D8-barD8SS}(b).  It was shown in \cite{SS} that this classical
configuration is stable, by analyzing small fluctuations around this
configuration and checking that the energy density associated with
them is non-negative.

For general $L$, one might apriori
think that there could be two ways for the D8-branes
to connect to the anti-D8-branes -- one in which they connect by curving
towards positive $x_4$ (in the direction where the D8-anti-D8 distance
is $L$) and the other by curving towards negative $x_4$ (in the direction
where the distance is $2\pi R - L$). However, smooth configurations only
occur when the integral (\ref{forl}) is less than or equal to $\pi R$, so there is
only one smooth configuration for every value of $L$.


Note that the Sakai-Sugimoto model has 3 dimensionful parameters : $\lambda_5$, $L$
and $R$, and gravity is reliable whenever $\lambda_5 \gg R$. The physics
depends on the two dimensionless ratios of these two parameters; we will
see that in the gravity limit
the mass of the (low-spin) mesons is related to $1/L$ while the mass of
the (low-spin) glueballs is related to $1/R$. As discussed above, in the limit
$\lambda_5 \ll R$ this theory approaches (large $N_c$) QCD at low energies.
This remains true also after adding the flavors, at least when $L$ is of
order $R$ (note that
the parameter $L$ 
is not visible in the four-dimensional low-energy effective
action). However, this QCD limit is opposite from the supergravity
limit, and there are no known reliable methods to analyze it.

\subsection{The spectrum of mesons}

The mesons of the Sakai-Sugimoto model are described by strings ending
on the D8-branes. It is convenient to separate the discussion to
low-spin mesons and high-spin mesons.  The former are described via
modes of the massless fields living on the D8-branes, whereas the
latter are associated with string configurations that ``fall'' from
the D8-branes down to the wall at $u=u_\Lambda$, stretch along the
wall and then go back up again \cite{KPSV}. 

Let us start with the low-spin mesons. These mesons correspond on the
 string theory side to the fluctuations of the massless fields on the probe
 branes.  The fluctuations of the gauge fields on the branes give
 pseudo-vector and scalar mesons as well as massless ``pions'', and the
 fluctuations of the scalar field describing the embedding of the
 branes give massive scalar mesons. In \cite{SS} these two sectors of
 the mesonic spectrum were computed for the special case of
 $u_0=u_\Lambda$, namely $L= \pi R$. It is not difficult to generalize
 this analysis to general values of $L$ which we are interested in
 here, using the analysis of the fluctuations performed in \cite{CPS}
 which we summarize below (the discussion in \cite{CPS} is for a
 non-critical $AdS_6$ model, but one can easily apply the same
 analysis for our model).

As an example we describe in detail
the modes coming from the components of the gauge
field living on the D8-branes which are not in the $S^4$
directions.
If we write the action using the coordinate $u$ (instead of using
$x_4$ as in (\ref{eightaction})), the equations of motion
for these fields that follow from the DBI action are given by\footnote{
We perform the analysis for a single D8-brane, but the generalization
to $N_f$ D8-branes is simply performed by turning all the fields into
$N_f\times N_f$ matrices, turning the derivatives into
$U(N_f)$-covariant
derivatives, and taking a trace in the action.}
\bea
u^{5/2}\gamma(u)^{1/2}\partial_\mu F^{\mu\nu}-\partial_u
(u^{5/2}\gamma(u)^{1/2}F^{\nu u})&=&0,
\label{eqngauge1}\\
\label{eqngauge2} \partial_\mu F^{\mu u}&=&0\, ,
\eea
where $\mu,\nu=0,1,2,3$ and where we have defined
\be
\gamma(u) \equiv \frac{u^8}{f(u) u^{8}-f(u_0) u_0^8}.
\ee
We now separate variables and expand the five dimensional gauge
fields in modes in the radial direction as follows :
\be
A_\mu(x^\nu, u) = \sum_n B_\mu^{(n)}(x^\nu)\psi_n(u), \qquad
A_u(x^\nu, u) = \sum_n \pi^{(n)}(x^\nu)\phi_n(u).
\ee
Using this decomposition, equation (\ref{eqngauge2}) reads:
\be
\label{geneom}
\sum_n\left( \tilde{m}_n^2\pi^{(n)}(x^\nu)\phi_{n}(u)-
  (\partial_\mu B_{(n)}^\mu(x^\nu))\partial_u\psi_{n}(u)   \right)=0,
\ee
where $\tilde{m}_n$ is the four-dimensional mass of the field
$\pi^{(n)}$, namely  $\eta^{\mu\nu}\partial_\mu \partial_\nu
\pi^{(n)}=\tilde{m}_n^2 \pi^{(n)}$. We can now choose a Lorentz gauge
$\partial_\mu A^{\mu} = 0 \to \partial_\mu B^\mu_{(n)}=0$, and in this
gauge it is easy to see that all the $\pi^{(n)}$
fields are set to zero by (\ref{geneom}) except for the Goldstone bosons
$\pi^{(0)}$
with $\tilde{m}_0=0$, which
survive as a massless field for any value of $u_0$.
The other equation of motion (\ref{eqngauge1}) fixes
the form of $\phi_{0}$ to be
\be
\phi_0\propto \frac{\gamma^{1/2}}{u^{5/2}}\,\,.
\label{phi(0)}
\ee
Finally, the eigenvalue equation for the modes $\psi_{n}(u)$ reads 
\be
-u^{1/2}\gamma^{-1/2} \partial_u (u^{5/2}  \gamma^{-1/2}
\partial_u \psi_{n}) =
R_{D4}^3 m_n^2 \psi_{n},
\label{vectoreq}
\ee
where  $m_n$ are  the   masses  of  the four-dimensional  vector  fields
$B^{(n)}_\mu$, such that their equation of motion is
$\eta^{\nu\rho}\partial_\nu        \partial_\rho
B^{(n)}_\mu={m}_n^2   B^{(n)}_\mu$. Imposing  appropriate
orthonormality  conditions \cite{CPS}, one  finds that the
effective four-dimensional action for  these fluctuations of
the gauge field living on the probe D8-branes can be written as:
\be\label{gaugeactionbis}
S=-\int d^4x {\rm tr}\left[ \frac12 \partial_\mu\pi^{(0)}
\partial^\mu\pi^{(0)}+\sum_{n\geq 1} \left( \frac14
F_{\mu\nu}^{(n)}F^{\mu\nu\;(n)}+\frac12
m_n^2 B_\mu^{(n)}B^{\mu\;(n)}\right)\right]\,\,.
\ee
The fields $\pi^{(0)}$ are the Goldstone bosons associated
with the spontaneous breaking of the $U(N_f)_L\times U(N_f)_R$ global
symmetry to the diagonal $U(N_f)$ (which is the global symmetry
of the remaining action (\ref{gaugeactionbis})).
As in \cite{SS}, one can also go to the $A_u=0$ gauge, where there are
no $\pi^{(n)}$ fields and the Goldstone bosons are encoded in the
boundary value (as $u\to \infty$) of $A_\mu$. Note that the pions
(the $\pi^{(0)}$ fields) are massless for
any value of $u_0$ (or for any value of $L$), as expected since the
quarks are massless (in the sense of having a vanishing current
algebra mass) for any value of $L$.

The explicit mass spectrum of the vector mesons is found by looking
for normalizable eigenfunctions of (\ref{vectoreq}). In the special
case of $u_0=u_{\Lambda}$ this spectrum was derived in \cite{SS}.  For
generic probe profiles, for which $u_0\neq u_\Lambda$, the
spectrum was computed in \cite{CPS} for a similar non-critical string
model. In general this spectrum is quite complicated.
In  the limit of $u_0 \gg u_{\Lambda}$ or $L \ll R$ the spectrum simplifies. In this
limit we can approximate $f(u)\simeq 1$, and there is a scaling
symmetry of (\ref{vectoreq}) which implies that all $m_n^2$ are
proportional to $u_0 / R_{D4}^3$. Equation (\ref{lzerotemp}) then
implies that this is proportional to $1/L^2$, so in this limit all the
low-spin meson masses are simply proportional to $1/L$. This
illustrates the separation of scales between the low-spin mesons and the
low-spin glueballs (whose masses are proportional to $1/R$ in the supergravity
approximation) in the Sakai-Sugimoto model (as emphasized in
\cite{AHJK}).

\begin{figure}[t]
\begin{center}
\vspace{3ex}
\includegraphics[width=.65\textwidth]{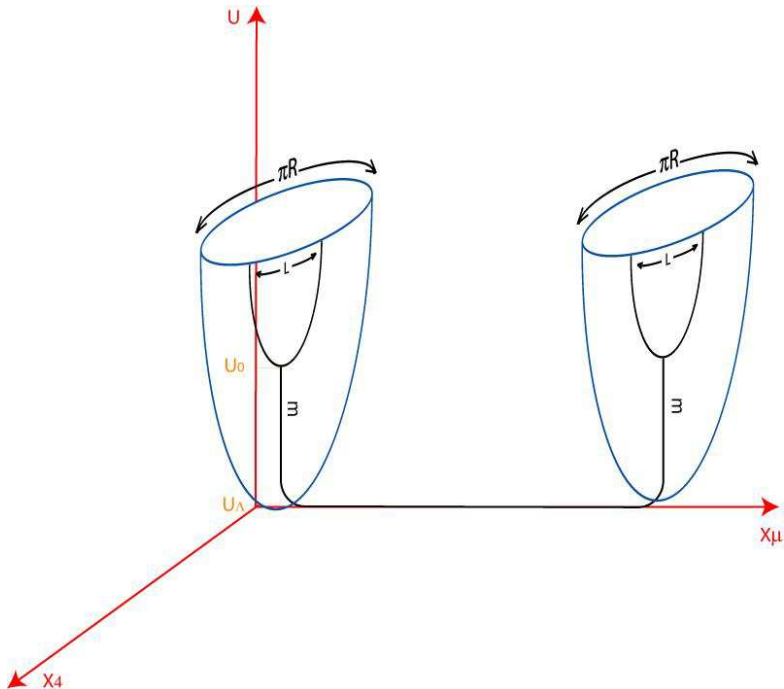}
\end{center}
\caption{The high-spin ``stringy'' meson is a string starting at the
lowest point of the probe branes $u=u_0$, going down to the wall,
stretching horizontally in space along the wall, and then going back up
vertically to the probe branes at $u=u_0$.
  \label{stringymeson}}
\end{figure}

The DBI action of the D8-branes only includes mesons of spin one or
less.  To describe higher-spin mesons we need to look at more general
string configurations that start and end on the probe branes.  For
large spin these strings are long and can be described
semi-classically.  The relevant string configurations can be
schematically decomposed into three parts : a segment from the
probe brane (at $u=u_0$) to the wall ($u=u_\Lambda$), then a segment
that stretches along the wall in the spatial directions, and then
another ``vertical'' part stretching from the wall back to the probe 
brane, as depicted in figure \ref{stringymeson}.

By computing the energy of the string it was shown in \cite{KPSV} that
this string is equivalent to a string in flat space-time with two
massive endpoints. The mass of the end-points, which can be identified
with the  constituent quark mass $m_q^C$, is given by
%
\cite{CPS}
\be\label{quarkmass}
m^C_q =\frac{1}{2\pi\alpha'}\int_{u_\Lambda}^{u_0}
\sqrt{-g_{tt}g_{uu}}du=
\frac{1}{2\pi\alpha'}\int_{u_\Lambda}^{u_0} f^{-\frac12}(u)du,
\ee
which is simply the mass of a string stretched from the D8-branes
to the minimal value of $u$. Note that for small $L$ this constituent
mass scales as $u_0/\alpha' \propto \lambda_5 / L^2$.
Let us emphasize again that the bare mass and current algebra masses
are always zero in our configurations, and there is no simple
description of the deformation associated with turning on a bare mass
for the quarks (which corresponds to turning on a non-normalizable
mode of the string connecting the D8-branes to the anti-D8-branes).
%
Using the map to the string with massive endpoints it is
straightforward to calculate the classical mass and angular momentum
of these stringy mesons \cite{KPSV,Wilczek}.  Somewhat surprisingly, it was found
in \cite{CPS} (in a similar model to the one we describe) that for
$u_0$ of order $u_\Lambda$ there is a range of values of
$u_0/u_\Lambda$ for which the masses of the light vector and pseudo
scalar mesons scale roughly linearly with $m^C_q$, reinforcing its
identification as a constituent quark mass.
The decay processes of the stringy mesons were discussed in \cite{PSZ}.
%
%
%
%



\newsection{Thermodynamics of the Sakai-Sugimoto model}

\subsection{Review of the bulk thermodynamics}

In the large $N_c$ limit (with finite $N_f$), the effect of the D8-branes
on the background is subleading; the difference in free energies between
different bulk backgrounds is of order $N_c^2$, while the contribution
of the D8-branes is of order $N_c*N_f$. Thus, we can analyze the bulk
thermodynamics independently of the presence of the D8-branes, and then
add them in as probes to the dominant bulk background at each
temperature.

\begin{figure}[t]
\begin{center}
\vspace{3ex}
\includegraphics[width=.85\textwidth]{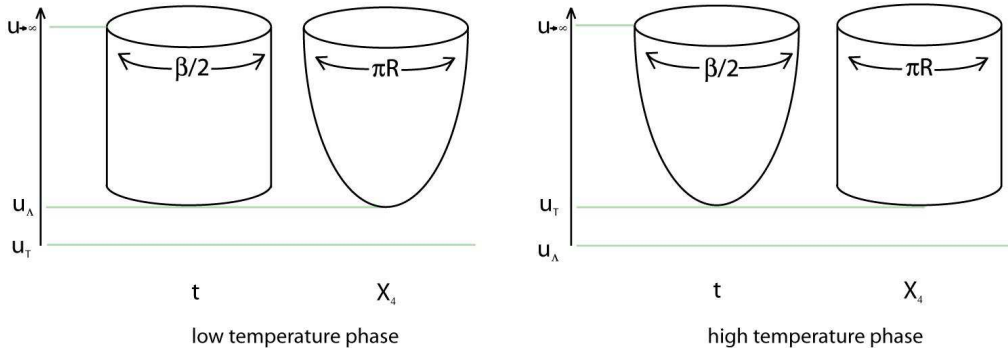}
\end{center}
\caption{ The topology of the solutions which dominate in the 
  low temperature (confined) and high temperature (deconfined) phases
  of the Sakai-Sugimoto model, as reflected in the $(t, u)$ and $( x_4, u)$
  submanifolds.
  \label{phases}}
\end{figure}

The thermodynamical behavior of the bulk is
analogous to the thermodynamical behavior of other theories
compactified on circles with anti-periodic boundary conditions
for fermions in the
gravity approximation (see,
e.g., \cite{Wpure,MalStro}). In the gravity
approximation, in order to study the theory at a finite temperature
$T$, we need to look at Euclidean gravitational backgrounds which are
asymptotically equal to (\ref{SSmodel}), but with the time direction
Euclidean and identified with a periodicity $\beta = 1 / T$, and with
anti-periodic boundary conditions for the fermions along this
direction (as well as along the $x_4$ direction). One such background
is obviously given by the Euclidean continuation of (\ref{SSmodel})
itself, with $t$ periodically identified; it is easy to verify that
this is a smooth background which is weakly curved (at least until the
temperature reaches a number of order the square root of the confining
string tension $T_{st} \simeq \lambda_5 / R^3$, at which stage the
size of the time circle
at the minimal radial position $u=u_{\Lambda}$
becomes of order the fundamental string scale
and a tachyon develops). In
this background the periodicity of $t$ is arbitrary and equal to
$\beta$, and the periodicity of $x_4$ is $2\pi R$ with $R$ related to
$u_{\Lambda}$ by (\ref{relru}).

In the background described above the $x_4$ circle shrinks when $u$
reaches its minimal value, but the $t$ circle never shrinks to zero
size.  Another background with the same asymptotics is given by
exchanging the behavior of the $t$ and $x_4$ circles, or equivalently
by moving the factor of $f(u)$ in (\ref{SSmodel}) from the $dx_4^2$
term to the $dt^2$ term. Now, the $t$ circle shrinks to zero size at
the minimal value of $u$ (which is now related to $\beta$ rather than
to $2\pi R$, and given by $u_T = 16 \pi^2 R_{D4}^3 / 9 \beta^2$), 
while the $x_4$ circle
never shrinks. This background also exists as a classical solution for
any value of the periodicities of $x_4$ and $t$ (and develops a
tachyon for very small values of $R$).

These two backgrounds are the only known smooth Euclidean solutions
which asymptotically approach (\ref{SSmodel}) \footnote{Note that
  asymptotically these backgrounds become strongly coupled so we
  should really lift them to M theory, and discuss the asymptotics in
  the M theory regime, but in the large $N_c$ limit the strong
  coupling region can be arbitrarily far away from the regions of
  finite $u$ which we are interested in so this does not play any
  role.}. Thus, in order to decide which background dominates at a
given temperature $T$ we need to compute the free energies of these
backgrounds, given (in the gravitational approximation) by the
classical supergravity action times the temperature, and see which one
has a lower free energy. As usual, the classical action actually
diverges and needs to be regulated. One way to regulate it is by
looking at the difference between the actions of the two solutions,
which turns out to be finite (this method is not always reliable, but
it gives the correct answer in this case). Another way is to add
counter-terms at a large $u$ cutoff that make the action finite as the
cutoff goes to infinity. In the first method it is clear that the free
energies of the two solutions agree when the (asymptotic)
circumferences of the two circles are equal, $\beta = 2 \pi R$, since
the two solutions are identical then except for a relabeling of the
coordinates. Thus, at this temperature $T_d = 1 / 2\pi R$ there is a
first order phase transition between the two backgrounds (the
transition is of first order since the solutions do not smoothly
connect there, but continue to exist as separate solutions both below
and above the transition). It is easy to see that the background in
which the $x_4$ circle shrinks to zero size dominates at low
temperatures $T < 1 / 2 \pi R$, while the background in which the $t$
circle shrinks dominates at high temperatures, $T > 1 / 2\pi R$;
intuitively, the circle which is smaller prefers to shrink to zero
size. The relevant computation is described in the appendix, and leads
(\ref{difffree}) to a difference of free energy densities proportional to
$N_c^2 R (g_s N_c) [(2\pi T)^6 - 1 / R^6]$.

The physical interpretation of this phase transition (which is a close
relative of the Hawking-Page transition in asymptotically AdS
backgrounds \cite{Wpure}) is straightforward. If we compute the
quark-anti-quark potential in the two backgrounds (before adding the
D8-branes) using the methods of \cite{Maldacena:1998im,Rey:1998ik,KSS},
 we find that in the low-temperature background in which $\sqrt{g_{tt}g_{xx}}$
is finite at the minimal value of $u$ the potential is linear in the
distance, corresponding to confinement, while in the high-temperature
background in which $\sqrt{g_{tt}g_{xx}}$ goes to zero at the minimal value of $u$
the potential decays with the distance, corresponding to a deconfined
phase. Similarly, a computation of the value of the free energy in the
two phases (which requires adding appropriate counter-terms) yields a
result of order $N_c^0$ in the low-temperature phase, and a result of
order $N_c^2$ in the high-temperature phase.

\begin{figure}[t]
\begin{center}
\vspace{3ex}
\includegraphics[width=.65\textwidth]{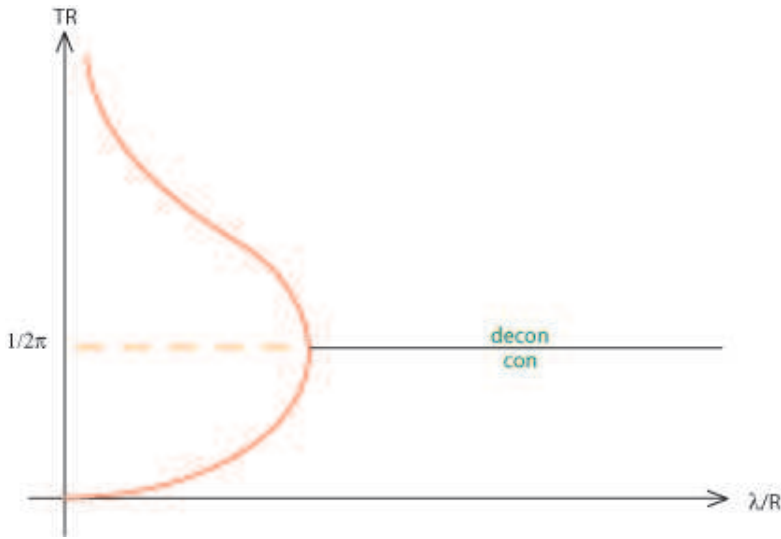}
\end{center}
\caption{A minimal conjecture for the schematic form of the
  phase diagram of the $4+1$
dimensional maximally supersymmetric $SU(N_c)$ gauge theory
compactified on a circle with anti-periodic boundary conditions for
the fermions, as a
function of the dimensionless coupling constant $\lambda_5/R$ and 
the dimensionless 
temperature $TR$, based
on the known results in the various limits and on the $T
\leftrightarrow 1/2\pi R$ symmetry. All solid lines in the diagram denote
phase transitions; the dashed line may or may not be a phase transition
line.
\label{condecon}}
\end{figure}

As mentioned above, as we increase $R$ compared to the scale set by
the five dimensional gauge coupling $\lambda_5$, the supergravity
background becomes highly curved, and the theory at low energies
approaches the four dimensional pure Yang-Mills theory. It is believed
(though it is difficult to prove) that the deconfinement transition
described above is connected to the deconfinement transition of the
large $N_c$ pure Yang-Mills theory in this limit. However, this
connection is somewhat subtle \cite{AMW}, since even for small
$\lambda_5/R$ it is clear by the symmetry arguments described above
that there is always either a phase transition line at $T=1/2\pi R$, or else
the dominant phase at this temperature must be invariant under the
exchange $T \leftrightarrow 1 / 2\pi R$ (which is not the case for the
confined and deconfined phases discussed above). On the other hand,
the deconfinement phase transition in the pure Yang-Mills theory
occurs at a temperature of order $\Lambda_{QCD} \ll 1/R$. Due to the
$T \leftrightarrow 1/2\pi R$ symmetry there must then be at least one
more phase transition line at a temperature much larger than
$1/R$. Thus, we see that new phases must appear as we approach the
four dimensional Yang-Mills limit. The phase structure of this theory
in that limit includes several phase transition lines, which,
presumably, all join into the line $T = 1 / 2\pi R$ in the gravity
limit. A minimal conjecture for the phase structure is depicted in
figure \ref{condecon}; note that with this conjecture the
deconfinement transition line in the supergravity limit is connected
to the deconfinement transition line in the four dimensional
Yang-Mills limit, but this connection does not seem to be
smooth\footnote{This conjectured phase diagram is the same as the
phase diagram of the same theory reduced on the three spatial
directions which was discussed in \cite{Aharony:2005ew}, suggesting
that perhaps the phase structure does not change when the three
infinite spatial dimensions are compactified on a torus. We thank M. Van
Raamsdonk for discussions of this point.}.

\subsection{The low temperature phase}

As described above, the background corresponding to the low-temperature
phase of the theory
is identical to that of the Sakai-Sugimoto model at zero temperature
(\ref{SSmodel}),
apart from the fact that the time direction is Euclidean and
compactified with a circumference $\beta=1/T$, where $T$ is the
temperature of the dual gauge theory.  
Hence, just as for the zero temperature case,
the dual gauge theory is in the confining phase and the string tension
is given by (\ref{stringauge}). Furthermore, the solution for the
profile of the D8 brane is still given by (\ref{D8profile}). Note that
this agrees with general arguments that the chiral condensate (in the
large $N_c$ limit) should be independent of the temperature in the
confined phase \cite{Neri:1983ic,Pisarski}. Defining $y\equiv u/u_0$,
$y_{\Lambda}\equiv u_{\Lambda}/u_0$, and $f(y)\equiv 1-(y_\Lambda/y)^3$
(such that $f(1)=1-y_\Lambda^3$), we find from (\ref{D8profile}) that
\be 
\label{nprofile}
u'=
\left( \frac{u}{R_{D4}}
\right)^{3/2}f(y)\sqrt{y^8\frac{f(y)}{f(1)}-1}. \ee 
Substituting this
outcome of the equation of motion, the action is now \bea
\label{lowtaction}
S_{DBI} &=&
2\frac{\hat T_8 R_{D4}^{3/2}u_0^{7/2}}{g_s}
\int_1^\infty dy \frac{y^{5/2}}{\sqrt{f(y)}} 
\frac{1}{\sqrt{1-\frac{f(1)}{f(y)}\frac{1}{y^8}}}\CR
&=&\frac{2 \hat T_8 R_{D4}^{3/2}u_0^{7/2}}{3 g_s} \int_0^1 dz
\frac{1}{z^{13/6} \sqrt{1- (1-y_\Lambda^3) z^{8/3} -y_{\Lambda}^3 z}}. \eea 
%

%
The form of the profile (\ref{nprofile}) of the D8-branes is drawn in figure
\ref{D8-barD8SS}(a). Again, the topology of the background forces a profile 
that smoothly connects the D8-branes to the anti-D8 branes, 
signaling chiral symmetry breaking in the
dual gauge theory.

The relation between $L$ and $u_0$ is precisely the same as for zero
temperature (\ref{lzerotemp}). For small values of $L$ we find that
the dependence of the action (\ref{lowtaction}), which is proportional
to the vacuum
energy of the system, on the separation distance is given by
\be S_{DBI}\propto
\frac{\hat T_8 R_{D4}^{3/2}}{g_s L^7}. \ee
For general values of $L$ the dependence is more complicated.

\subsection{The high temperature phase}

The gravity background associated with the high temperature phase of
the dual gauge theory takes the form 
\bea \label{actionhigh}
ds^2&=&\left(
\frac{u}{R_{D4}} \right)^{3/2}\left [ f(u) dt^2+ \delta_{ij}dx^{i}dx^j
+ dx_4^2\right ] +\left( \frac{R_{D4}}{u} \right)^{3/2} \left [ u^2
d\Omega_4^2 + \frac{du^2}{f(u)} \right ],
\label{unflavmetr}\CR
F_{(4)}&=& \frac{2\pi N_c}{V_4}\epsilon_4, \quad  
e^\phi = g_s\left( \frac{u}{R_{D4}} \right)^{3/4},\quad
 R_{D4}^3 \equiv \pi g_s N_c l_s^3,\quad 
f(u)\equiv 1-\left( \frac{u_T}{u} \right)^3,
\eea
in the same notations as in (\ref{SSmodel}).
The time direction $t$ now shrinks to zero size at the minimal value of $u$,
$u=u_T$, and in order to avoid a singularity there 
the time direction must have a period of
\be
\label{betadef}
\delta t = \frac{4\pi}{3}\left( \frac{R_{D4}^3}{u_T} \right)^{1/2} = \beta
\ee 
which we identify with the inverse temperature of the system $\beta = 1/T$.
On the other hand, in this solution the $x_4$ direction 
can have an arbitrary periodicity $2\pi R$.

We expect that again the D8 probe branes span the coordinates $t, x^i, 
\Omega_4$, and are described by some curve $u(x_4)$. 
The induced metric on the D8 branes now reads
\bea
ds^2_{D8}&=&\left( \frac{u}{R_{D4}} \right)^{3/2}\left [ f(u) dt^2+  \delta_{ij}dx^{i}dx^j \right ] 
+\left( \frac{u}{R_{D4}} \right)^{3/2} \left [1+ \left( \frac{R_{D4}}{u} \right)^{3} \frac{{u'}^2}{f(u)}\right ]dx_4^2 \CR
 &+&
 \left( \frac{R_{D4}}{u} \right)^{3/2} u^2 d\Omega_4^2  
\eea  
where $u'=du/dx_4$.
Just as for the low temperature phase, here too the CS term does not
affect the solution of the equations of motion.  Upon substituting the
determinant of the induced metric and the dilaton, the DBI action is
given by
\be
S_{DBI} = T_8 \int dt d^3 x dx_4 d^4\Omega e^{-\phi} \sqrt{-\det(\hat g)}
        =  {\hat T_8 \over g_s} \int  dx_4 u^4\sqrt{f(u)} \sqrt{1+ 
\left( \frac{R_{D4}}{u} \right)^{3} \frac{{u'}^2}{f(u)}},
\label{newaction}
\ee
where $\hat T_8$ includes the outcome of the integration over all the
coordinates apart from $dx_4$.

\begin{figure}[t]
\begin{center}
\vspace{3ex}
\includegraphics[width=.75\textwidth]{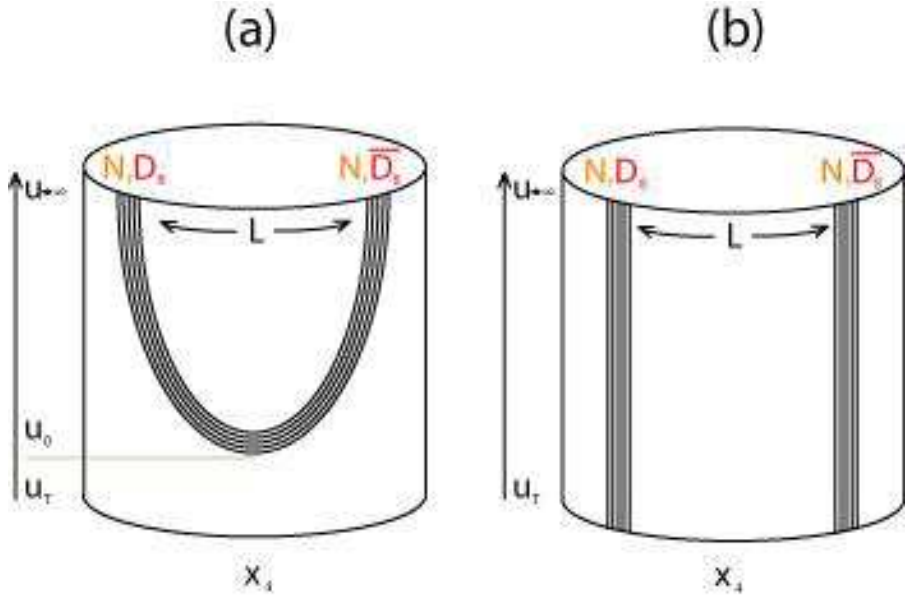}
\end{center}
\caption{The possible configurations of the D8 and anti-D8 probe
  branes in the high temperature (deconfined) phase.  A generic connected
  configuration with an asymptotic separation of $L$, that stretches
  down to a minimum at $u=u_0$, is drawn in (a). 
This corresponds to
  the chiral symmetry broken phase. Figure (b) depicts a
  chiral-invariant solution, with separate D8 and anti-D8-branes.
  \label{D8-barD8}}
\end{figure}

Again, conservation of the Hamiltonian of (\ref{newaction}) implies that
\be
\frac{\sqrt{f(u)}u^4}{\sqrt{1+ \left( \frac{R_{D4}}{u}
\right)^{3} \frac{{u'}^2}{f(u)}}}= {\rm constant} =
u_0^4\sqrt{f(u_0)},
\ee 
where on the right-hand side of the equation we assumed
that there is a point $u_0$ where the profile of the brane has a
minimum, $u'(u=u_0)=0$.  
Defining again $y\equiv u/u_0$, we find that 
\be u'= \left( \frac{u}{R_{D4}}
\right)^{3/2}\sqrt{f(y)} \sqrt{\frac{f(y)}{f(1)}y^8-1}.
\ee 
Substituting the outcome of the equation of motion, the action is
now 
\be S_{DBI} ={2\hat T_8 R_{D4}^{3/2}u_0^{7/2}\over g_s} \int_1^\infty dy
\frac{y^{5/2}}{\sqrt{1-\frac{f(1)}{f(y) y^8}}
}. \ee 
Since we assumed $u'(u=u_0)=0$,
this action corresponds to a configuration where the
D8-branes merge smoothly into the anti-D8-branes,
as in figure \ref{D8-barD8}(a).

As discussed above, 
this configuration exhibits in the string theory dual
the phenomenon of $\chi$SB in the gauge theory.
However, there is yet another classical configuration that solves the
equations of motion. This is a configuration of a constant $x_4$, $x_4(u)=0$
and $x_4(u)=L$ (or
$u'\rightarrow \infty$), see figure \ref{D8-barD8}(b). 
This configuration is possible
in the high-temperature phase since in this phase the $t$ circle shrinks
to zero at $u=u_T$, so the D8-brane can just smoothly end there (and
wrap around the ``cigar'' in the $t-u$ plane); in the Minkowski space
continuation of this solution, the D8-brane would go into the horizon
of the black D4-branes. For such a configuration that
naturally describes the situation of an unbroken chiral symmetry, the
action reads (summing over the D8 and anti-D8 contributions)
\be S^0_{DBI} = {2 \hat T_8 \over g_s} \int_{u_T}^{\infty} du u^4 \left(
\frac{R_{D4}}{u} \right)^{3/2} = {2 \hat T_8 R_{D4}^{3/2}
u_0^{7/2}\over g_s} \left
[\int_1^\infty dy y^{5/2} +\int_{y_T}^1 dy y^{5/2} \right],
\ee 
where
$y_T\equiv u_T/u_0$.

We are now facing the question of which of the two possible
configurations is preferred.  This is determined
by the free energy, or the action, of the configurations. The relevant
quantity is the following difference between the two actions
\bea \Delta S &\equiv& \frac{g_s(S-S^0)}{2 \hat T_8 R_{D4}^{3/2}
u_0^{7/2}} = \left
\{\int_1^\infty dy y^{5/2} \left [
\frac{1}{\sqrt{1-\frac{f(1)}{f(y)}y^{-8}}}-1\right ] - \int_{y_T}^1 dy
y^{5/2} \right \} \CR &=& \frac{1}{3}\left
\{\int_0^1 dz \frac{1}{z^{13/6}}\left [ \frac{\sqrt{1-y_T^3 z}}
{\sqrt{1-y_T^3 z-(1-y_T^3)z^{8/3}}}-1\right ]\right \} -\frac{2}{7} (
1-y_T^{7/2}). \eea

\begin{figure}[t]
\begin{center}
\vspace{3ex}
\includegraphics[width=.65\textwidth]{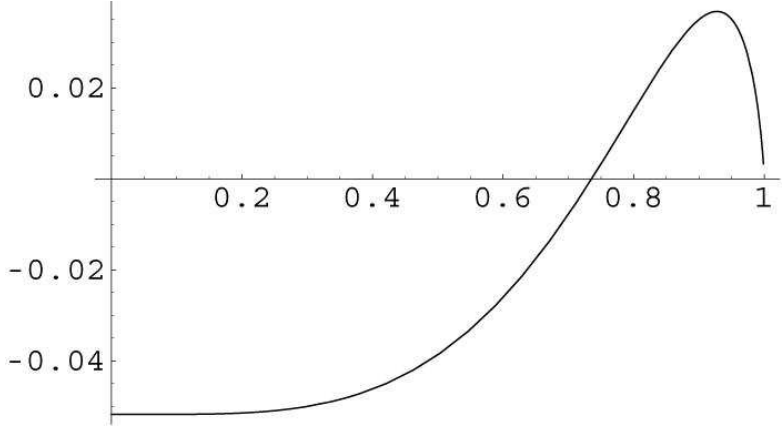}
\end{center}
\caption{ $\Delta S$ as a function of $y_T$, in units of
$2{\hat T_8} R_{D4}^{3/2}u_0^{7/2}/g_s$. \label{deltas}}
\end{figure}

We are unable to evaluate $\Delta S$ analytically so we do it
numerically. The result is drawn in figure \ref{deltas}, which shows $\Delta S$
as a function of $y_T$.  It is easy to see from this figure that for
$y_T>y_T^c\sim 0.73572$, $\Delta S$ is positive and hence the preferred
configuration is that of disconnected D8 and anti-D8-branes, 
whereas in the region where $y_T<y_T^c\sim 0.73572$ the system will
be dominated by a smooth configuration that connects the D8 and anti D8
branes, namely the chiral symmetry broken phase.  

Next, we would like
to express the critical point in terms of physical quantities.
Recall that $y_T\equiv u_T/u_0$. $u_T$ is related to the
temperature in the gauge theory by (\ref{betadef}).
$u_0$ is the minimal point of
the connected probe brane configuration, which can be related to the asymptotic
distance $L$ between the positions of the D8 and anti-D8-branes (see
figure \ref{D8-barD8}) in the following manner\footnote{
In fact, as we discussed above, this integral could equal either $L$
or $2\pi R - L$, but it is easy to see that the second option is always
sub-dominant wherever it exists.}
\bea 
\label{lcritical}
L &=& \int dx_4 = 2 \int_{u_0}^{\infty}
\frac{du}{u'}= 2 \left( \frac{R^3_{D4}}{u_0} \right)^{1/2}\int_1^\infty
dy \frac{y^{-3/2}}{\sqrt{f(y)} \sqrt{\frac{f(y)}{f(1)}y^8-1}}
\CR &=& \frac{2}{3} \left( \frac{R^3_{D4}}{u_0}
\right)^{1/2}\sqrt{1-y_T^3}\int_0^1 dz
\frac{z^{1/2}}{\sqrt{1-y_T^3 z}\sqrt{1-y_T^3 z-(1-y_T^3)z^{8/3}}}. \eea
Note that for small $L$, namely for $y_T<<1$, we find $u_0 \propto
R^3_{D4}/L^2$ and again $S\propto 1/L^7$. We would like to
relate the asymptotic distance $L$ at the transition to the temperature
of the transition. 
%
%
Evaluating the integral in (\ref{lcritical}) at the transition
temperature $y_T^c\simeq 0.73572$ gives $L_c \simeq 0.751 (R_{D4}^3
/ u_0)^{1/2}$, and using equation (\ref{betadef}) we find
\be 
\label{crittemp}
T_{\chi SB} \simeq 0.154 / L.
\ee

\begin{figure}[t]
\begin{center}
\vspace{3ex}
\includegraphics[width=.65\textwidth]{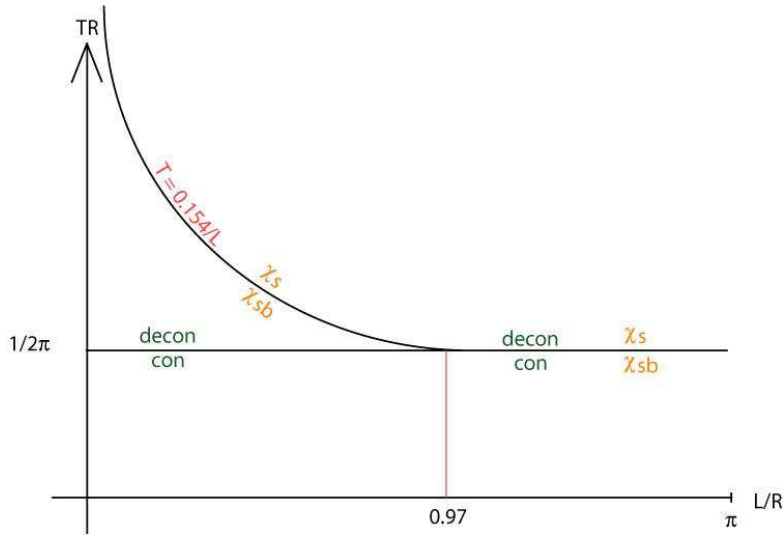}
\end{center}
\caption{The phase diagram of the Sakai-Sugimoto model at finite
  temperature, in the gravity approximation $\lambda_5 \gg R$. In
  this approximation the phase structure depends only on the two
  dimensionless parameters $TR$ and $L/R$. For small $L/R$ the
  deconfinement and chiral symmetry restoration transitions are separate, while
for $L/R>0.97$ there is a single transition.
\label{phasediagram}
}
\end{figure}

Recall that the deconfinement phase transition is at $T_d = 1 / 2\pi
R$, and that this analysis applies only in the deconfined phase.  For
$L > 0.97R$ the temperature in the deconfined phase is always higher
than (\ref{crittemp}), so the deconfinement and chiral symmetry
restoration transitions happen together. On the other hand, for $L <
0.97 R$, the two phase transitions are separate, and there is a finite
range between $T_d$ and $T_{\chi SB}$ for which the theory is
deconfined but with a broken chiral symmetry.
The full
phase diagram of the theory (in the supergravity approximation)
is drawn in figure \ref{phasediagram}.
Note that even though for the gravity approximation to be good
everywhere in space we required $\lambda_5 \gg R$, for
gravity to be good near the 8-branes it is actually enough to require
$\lambda_5 \gg L$, and at high temperatures gravity gives a
good approximation everywhere whenever $\lambda_5 \gg 1/T$.
Thus, we can trust our analysis of the phase transition whenever
$\lambda_5 \gg L$.
In particular, in the
$R\to \infty$ limit discussed in \cite{AHJK}, the theory is always
deconfined, and passes through a chiral symmetry restoration
transition at $T = T_{\chi SB}$ given by (\ref{crittemp}).
\footnote{Note added : in parallel with this paper, the paper
\cite{Parnachev:2006dn} appeared which discusses the same system in the
$R\to \infty$ limit, with identical results to ours.}

\subsection{The spectrum in the various phases}

In this subsection we discuss the spectrum of quark
configurations in the different phases described above; such a
discussion is more appropriate in Minkowski space, in the
finite-energy-density configurations (in the microcanonical ensemble)
obtained by Wick-rotating the Euclidean configurations discussed
above. In general there are two types of possible configurations. We
can have strings with both ends on D8-branes, corresponding to mesons
(for light strings these are described by the DBI action, and for large
spin they can be described by semi-classical strings, see figure
\ref{stringymeson}). And, if the
background contains a horizon (as in the deconfined phase) we can have
a string starting on an 8-brane and going into the horizon, corresponding
to a deconfined quark (or anti-quark).

As discussed above, the behavior in the low-temperature phase is very
similar to the behavior at zero temperature, as expected in a theory
with a small number of weakly interacting
massless particles (the ``pions''). There is a
discrete spectrum of mesons, and there are no free quark
states. Computing the spectrum of mesons (including the pseudo
vector and scalar mesons described in \S2.2) and glueballs in this phase
gives exactly the same results as for zero temperature.
The high spin mesons are described as classical string
configurations stretched from the D8 probe brane down to the ``wall''
at $u=u_{\Lambda}$
and back to the probe branes \cite{PSZ} (see figure
\ref{stringymeson}), and again there is no
difference between the zero temperature and low temperature results.

As we raise the temperature, at $T = 1 / 2\pi R$ we go over to the
deconfined bulk phase; as we saw, depending on $L/R$, we can either
get a phase with broken chiral symmetry or with unbroken chiral symmetry.
In the case of small $L/R$ with broken chiral symmetry
the D8-branes are still connected. In this phase the spectrum of
bulk modes has no mass gap, due to the presence of the horizon in
(\ref{unflavmetr}). On the other hand, a computation of the meson
spectrum using the D8-brane action still gives a discrete
spectrum as in \cite{SS}. However, unlike the low-temperature case,
here the spectrum of quark-anti-quark configurations is continuous
at high enough energies. This is because this spectrum includes
strings starting on the D8-brane and going down to the horizon $u=u_T$
(in Minkowski space such configurations are just deconfined quarks, while
in Euclidean space these strings must go back to the D8-brane, but they
can stretch any distance in the three dimensional space before doing so
with no cost in energy). These excitations give rise to a
continuous spectrum of quark-anti-quark configurations,
which starts above a mass gap which equals twice the
mass of a string stretching from $u_0$ to $u_T$,
$M_{gap} = 2 \int_{u_T}^{u_0} \sqrt{1 - (u_T / u)^3} du / l_s^2$.
This is obviously the analog of $2m_q^C$ defined in (\ref{quarkmass}),
and this also scales as $\lambda_5/L^2$ for small $L/R$.
Note that after we put in the D8-branes we can no longer use
the spatial Wilson loop as a diagnostic for confinement, since this can
be screened by the dynamical quarks, but the 
scaling of the free energy at large $N_c$ can still be used
to distinguish the confining phase at $T < 1 / 2\pi R$ from the
deconfined phase at $T > 1 / 2\pi R$.

Even when $L/R$ is small and such an intermediate phase exists, at high
enough temperature we always end up in a deconfined phase with a
restored chiral symmetry, in which the 8-branes end at the horizon
$u=u_T$. In this phase there is no longer a mass gap for
quark-anti-quark configurations, since strings stretched between the
8-branes and the horizon can be as light as we want, and the spectrum of
quark-anti-quark configurations is continuous (just like the bulk
spectrum).

\newsection{Summary}

In this paper we computed the phase structure of the Sakai-Sugimoto
model, which is an example of a confining gauge theory with chiral
symmetry breaking, related by a continuous change of parameters to
large $N_c$ QCD.  Our results are summarized in figure
\ref{phasediagram}.  We found that the topology of the solutions (in
the supergravity approximation) implies that the confined phase must
always break the chiral symmetry. On the other hand, in the deconfined
phase a computation is required to see whether chiral symmetry is
restored at the deconfinement temperature or above it, and we found
that in some range of parameters the first possibility is realized,
while in another range the second possibility occurs. When $L$ is much
smaller than $R$ 
we found that the chiral symmetry restoration scale is much higher than
the deconfinement scale, as expected because of the separation between
the confinement and chiral symmetry breaking scales in this case (as
emphasized in \cite{AHJK}). In this case the chiral symmetry
restoration happens at a scale related to the mass scale of the mesons
$1/L$, while the deconfinement transition happens at a scale related
to the glueball mass scale $1/R$.

The fact that confinement implies chiral symmetry breaking is true
also in a larger class of holographic duals of confining theories.
When the confinement is realized geometrically (this is not always
true \cite{PS}) there is a minimal value of the radial coordinate
$u=u_{\Lambda}$, and some spatial cycle shrinks there.  A chiral
symmetry should be realized by having two sets of branes (each
carrying one $U(N_f)$ factor), and chiral symmetry breaking should be
realized by the branes becoming connected. If the branes can end
without reconnecting, confinement does not imply $\chi$SB; clearly
this can only happen if the branes are wrapped on the cycle which
vanishes at the minimal value $u=u_{\Lambda}$.  Conversely, if the
branes are not wrapped around this cycle (as is the case in the
Sakai-Sugimoto model which we discussed here), they do not have
anywhere to end, so the chiral symmetry must be broken in the confined
phase. It would be interesting to relate this discussion to the
gauge theory discussions of chiral symmetry breaking in large $N_c$
gauge theories \cite{Neri:1983ic,Pisarski} which show that chiral symmetry
must be broken in the confined phase in the large $N_c$ limit.

In this paper we discussed the phase structure of the theory only in
the gravity regime $\lambda_5 \gg L$. It was recently realized in
\cite{AHJK} that there is another regime in which the theory is under
control, which is the weak-coupling regime $\lambda_5 \ll L$ with
$R$ going to infinity. It would be interesting to study the phase
structure also in this regime. There are also many other possible
generalizations of the Sakai-Sugimoto model which can be analyzed by
similar methods to the ones we used here, such as having several
different stacks of D8-branes (and of anti-D8-branes) separated in
the $x_4$ direction.

It is interesting to compare our results with the situation in large
$N_c$ QCD, which is related to our theory by taking the limit of small
$\lambda_5/R$. The technology to study large $N_c$ QCD with chiral
fermions on the lattice has only recently been developed, and
preliminary results presented in \cite{Narayanan:2005en}
suggest that in this
theory chiral symmetry restoration occurs at the same temperature as
the deconfinement transition, but that if one supercools the
deconfined phase below the deconfinement temperature one encounters a
critical temperature where the chiral symmetry is broken. This is
similar to what we find in the case of $L/R > 0.97$ (where we also
find a transition in the supercooled deconfined phase at the temperature
$T_c=0.154/L$), so it is plausible that the phase diagram in this
regime connects continuously to that of large $N_c$ QCD (up to the
caveats discussed at the end of \S3.1).


\section*{Acknowledgements}

We would like to thank S. Minwalla, H. Neuberger, B. Svetitsky, M. Van
Raamsdonk and especially
D. Kutasov for useful discussions.  J.S. would like to thank
O. Sonnenschein for helping with the drawings.  This work was
supported in part by a grant of DIP (H.52) and by the Albert Einstein Minerva
Center for Theoretical Physics at the Weizmann Institute. The work of
O.A. was supported in part by the Israel-U.S.  Binational Science
Foundation, by the Israel Science Foundation (grant number 1399/04),
by the Braun-Roger-Siegl foundation, by the European network
HPRN-CT-2000-00122, by a grant from the G.I.F., the German-Israeli
Foundation for Scientific Research and Development, and by Minerva.
The work of J.S. and S.Y. was supported in part by the Israel Science
Foundation (grant number 03200306).

\begin{appendix}

\section{Appendix: The bulk free energies of the low and high temperature phases}

In this appendix we present the calculation of the difference between
the free energies of the two possible solutions for the finite
temperature description of the bulk, described in section 3.1.  A
similar calculation in the context of the $AdS_5\times S^5$ model was
done in \cite{Wpure}. Rather than computing the action directly, we
translate our results into the notations of \cite{Kuperstein} and
use the results for the action computed there. Thus, we
parameterize the Euclidean metrics (\ref{SSmodel}) and
(\ref{actionhigh}) that we are interested in as
\begin{equation}       \label{stringmetric}
l_s^{-2}ds^2 = d\tau^2 +  e^{2\lambda(\tau)} dx_{\|}^2 
+ e^{2\tilde\lambda(\tau)} dx_{c}^2
 +e^{2\nu(\tau)} d\Omega_4^2,
\end{equation}
where $x_c$ is either $x_4$ or the Euclidean time (the one whose
circle shrinks to zero size at the minimal value of $u$), $x_{\|}$ are
the other four coordinates of $R^{4,1}$ (one of which is also
compactified), $\tau$ is the radial direction and $d\Omega_4^2$ is the
metric on the transverse $S^4$.  The background includes in addition
the dilaton $\phi$ and a four-form RR field.  For such backgrounds
that depend only on a radial direction, the full type IIA supergravity
action reduces to the following $0+1$-dimensional action
\cite{Klebanov:1998yy} :
\bea \label{actionrho} S = V \int d\rho \left(
-4(\lambda')^2 -(\tilde\lambda')^2-4(\nu')^2 +(\varphi')^2 + 12
e^{-2\nu-2 \varphi} -Q_c^2 e^{4 \lambda +{\tilde \lambda}-4 \nu - \varphi}
\right), \eea 
where the new radial coordinate $\rho$ is defined by $d\tau=
-e^{-\varphi}d\rho$, primes denote derivatives with respect to $\rho$,
and $ \varphi \equiv 2\phi-4\lambda-\tilde\lambda -4\nu $. $V$ is the
volume of all other directions (except $\rho$) in string units; this
includes the volume $V_3$ in the three spatial directions, the volume of the
two circles which is $2\pi R \beta / l_s^2$ and the volume of a unit
$S^4$ which is $2\pi^2$. The last term in the action corresponds to
the contribution of the RR flux, with $Q_c\equiv 3 \pi N_c / \sqrt{2}$
(see, for instance, \cite{Itzhaki:1998dd}) where $N_c$ is the
quantized RR flux.

The equations of motion associated with the action
(\ref{actionrho}) are
\bea
\lambda''-\frac12 Q_c^2 e^{2(4\lambda + \tilde \lambda-\phi)} &=& 0, \CR
\tilde\lambda''-\frac12 Q_c^2 e^{2(4\lambda + \tilde \lambda-\phi)} &=& 0, \CR
\phi''-\frac12 Q_c^2 e^{2(4\lambda + \tilde \lambda-\phi)} &=& 0, \CR
\nu''-3 e^{2(4\lambda +\tilde\lambda+3\nu-2\phi)} +
\frac12 Q_c^2 e^{2(4\lambda + \tilde \lambda-\phi)} &=& 0.
\eea
 It is straightforward to check that the following are the solutions
 of these equations \cite{Kuperstein,KS2} corresponding to the background
 (\ref{SSmodel}) :
\bea
\label{eomsol}
\lambda &=& -\frac{1}{4} \ln \left(\sinh(-4 b\rho) \right) 
+\frac{1}{4} \ln({u_{\Lambda}^3 \over {2 R_{D4}^3}})
- b\rho,
\qquad
\tilde{\lambda} = \lambda + 4 b\rho,
\CR
\phi &=& \lambda + \ln(g_s), \qquad \qquad \qquad \qquad \qquad \qquad
\qquad
\nu= \frac{\lambda}3 + \ln(R_{D4}/l_s),
\eea
where 
$b \equiv - Q_c u_{\Lambda}^3 / 4 \sqrt{2} R_{D4}^3 g_s$. 
Here we wrote the solution of the low temperature phase,
for the high temperature phase we have to replace $u_\Lambda$ with $u_T$.

Since the backgrounds (\ref{SSmodel}) and (\ref{actionhigh}) are
expressed in terms of the radial coordinate $u$, it is useful to
rewrite the action (\ref{actionrho}) in terms of integrals over $u$, which run from
$u_\Lambda$ to $\infty$ for the low temperature phase and from $u_T$
to $\infty$ for the high temperature phase. The action
(\ref{actionrho}) then reads (with dots denoting derivatives with
respect to $u$)
\be
\label{actionu} 
S = -V\int_{u_{\Lambda}}^{\infty} du \left\{ \left[-4{\dot\lambda}^2
-{\dot{\tilde\lambda}}^2-4{\dot\nu}^2
+{\dot{\varphi}}^2\right]\frac{du}{d\rho}
+\left[ 12 e^{8\lambda+2\tilde\lambda +6\nu-4\phi} -Q_c^2 e^{8 \lambda
+ 2 \tilde \lambda - 2\phi} \right]\frac{d\rho}{du} \right\}, \ee
where the minus sign arises because $d\rho/du$ is negative.
 Substituting the solution
(\ref{SSmodel}), the action takes the form
\be
\label{nactionu} S = -V\int_{u_{\Lambda}}^{\infty} du 
\left\{ \left[ \frac{15}{2u^2} \left(1+
  \frac{1}{[(\frac{u}{u_\Lambda})^3 -1]} \right)
\right]\frac{du}{d\rho} +\left[\left(\frac{12}{g_s^4 l_s^6} -
\frac{Q_c^2}{g_s^2 R_{D4}^6} \right)u^6 f(u) \right]\frac{d\rho}{du}
\right\}. \ee
%
At this point we need to compute the derivative $\frac{d\rho}{du}$
and its inverse. 
Using the relation  $e^{2\tilde\lambda}/e^{2\lambda}=
 [1-(\frac{u_\Lambda}{u})^3]$ and (\ref{eomsol})
 we find that $\frac{d\rho}{du}$ is given by
\be
\frac{d\rho}{du}= \frac{3}{8bu}\frac{1}{[(\frac{u}{u_\Lambda})^3-1]}.
\ee
Substituting this expression into the action we find that the divergence
at large $u$ is independent of $u_{\Lambda}$, so it makes sense to
subtract the expressions with $u_{\Lambda}$ and with $u_T$ and to obtain
a finite answer. The result
for the difference between the 
action densities in the low temperature 
phase and in the high temperature phase is given by (defining
${\hat b} = b l_s^3 / u_{\Lambda}^3$ which is a constant 
independent of $u_{\Lambda}$)
\bea 
\label{difffree}
{\Delta S \over V_3} \equiv {S_{low}-S_{high} \over V_3} &=&
\frac{4\pi^3 R \beta}{l_s^8} \left [ \frac{20\hat b}{3} +
\frac{1}{4\hat b}\left(\frac{6}{g_s^4}-\frac{Q_c^2}{2g_s^2
R_{D4}^6}\right )\right] (u_\Lambda^3-u_T^3) \CR &=& \frac{4\pi^3 R
\beta}{l_s^8} \frac{5}{g_s^2} (u_T^3-u_\Lambda^3) \CR &=& 20 \beta R
l_s (\frac49)^3\pi^6(g_s N_c) N_c^2 \left [ -\frac{1}{R^6}+
\frac{1}{(\frac{\beta}{2\pi})^6} \right ]\CR &=& 20 \beta R l_s
(\frac49)^3\pi^6(g_s N_c) N_c^2 [(2\pi T)^6 - M_{gb}^6].  \eea
%
%
In the third and fourth lines the difference of the actions was
expressed in terms of the radii of the circles and in terms of the
glueball mass scale and the temperature, respectively. The difference
of actions scales as $N_c^2$ in the 't Hooft large $N_c$ limit, as
expected.

The difference (\ref{difffree}) 
between the action densities is the same, up to
a factor of $\beta$, as the difference between the free energy
densities. It is now easy to see that for low temperatures,
$2\pi R < \beta$, the background (\ref{SSmodel}) has a lower
free energy so it dominates, while the opposite is the case for high
temperatures, as we claimed in section 3.1.



\end{appendix}

\end{document}